\documentclass[aps,prl,showpacs,superscriptaddress,prbib, twocolumn]{revtex4}
\usepackage{graphicx}
\usepackage{float}
\newcommand{\Fig}[1]{Fig.~\ref{#1}}

\begin{document}

\author{Clifford W. Hicks}
\affiliation{Stanford Institute for Materials and Energy Sciences, SLAC National Accelerator Laboratory, 2575
Sand Hill Road, Menlo Park, California 94025, USA}
\affiliation{School of Physics and Astronomy, University of St Andrews, St Andrews KY16 9SS, United Kingdom}

\author{John R. Kirtley}
\affiliation{Geballe Laboratory for Advanced Materials, Stanford University, Stanford, California 94305, USA}

\author{Thomas M. Lippman}
\affiliation{Stanford Institute for Materials and Energy Sciences, SLAC National Accelerator Laboratory, 2575
Sand Hill Road, Menlo Park, California 94025, USA}
\affiliation{Geballe Laboratory for Advanced Materials, Stanford University, Stanford, California 94305, USA}

\author{Nicholas C. Koshnick}
\affiliation{Geballe Laboratory for Advanced Materials, Stanford University, Stanford, California 94305, USA}

\author{Martin E. Huber}
\affiliation{Departments of Physics and Electrical Engineering, University of Colorado Denver, Denver,
Colorado, 80217, USA}

\author{Yoshiteru Maeno}
\affiliation{Department of Physics, Graduate School of Science, Kyoto University, Kyoto 606-8502, Japan}

\author{William M. Yuhasz}
\affiliation{Division of Materials Sciences and Engineering, Ames Laboratory, Ames, Iowa 50011, USA}

\author{M. Brian Maple}
\affiliation{Department of Physics, University of California-- San Diego, La Jolla, California 92093, USA}

\author{Kathryn A. Moler}
\affiliation{Stanford Institute for Materials and Energy Sciences, SLAC National Accelerator Laboratory, 2575
Sand Hill Road, Menlo Park, California 94025, USA}
\affiliation{Geballe Laboratory for Advanced Materials, Stanford University, Stanford, California 94305, USA}

\title{Limits on Superconductivity-Related Magnetization in Sr$_2$RuO$_4$ and PrOs$_4$Sb$_{12}$ from
Scanning SQUID Microscopy}

\date{October 2009}

\maketitle

{\sffamily We present scanning SQUID microscopy data on the superconductors Sr$_2$RuO$_4$ ($T_c=1.5$~K) and
PrOs$_4$Sb$_{12}$ ($T_c=1.8$~K). In both of these materials, superconductivity-related time-reversal
symmetry-breaking fields have been observed by muon spin rotation; our aim was
to visualize the structure of these fields. However in neither Sr$_2$RuO$_4$ nor PrOs$_4$Sb$_{12}$ do we observe
spontaneous superconductivity-related magnetization. In Sr$_2$RuO$_4$, many experimental results have been
interpreted on the basis of a $p_x \pm ip_y$ superconducting order parameter. This order parameter is expected
to give spontaneous magnetic induction at sample edges and order parameter domain walls. Supposing large
domains, our data restrict domain wall and edge fields to no more than $\sim$0.1\% and $\sim$0.2\% of the
expected magnitude, respectively. Alternatively, if the magnetization is of the expected order, the typical
domain size is limited to $\sim$30~nm for random domains, or $\sim$500~nm for periodic domains.}
\\

\noindent{ \large \bf \sffamily Introduction}
\\

Sr$_2$RuO$_4$ is a highly two-dimensional, layered perovskite superconductor with $T_c \approx 1.5~K$ in the
clean limit.  Its in-plane coherence length is $\xi_{ab}(0)=66$~nm~\cite{SROreview}, and magnetic penetration
depth $\lambda_{ab}=160$--190~nm~\cite{Baker09, Riseman98, Luke00} On the Fermi sheet thought to be 
dominant for superconductivity (the $\gamma$-sheet) the electron mass is $16m_e$ (5.5 times the band
mass)~\cite{SROreview}. PrOs$_4$Sb$_{12}$ is a heavy fermion superconductor of cubic symmetry, with $m^*$ in
the range of tens of $m_e$~\cite{Bauer02, Maple02}. Measurements of specific heat show an unusual double
superconducting transition, with $T_{c1} \approx 1.86$~K and $T_{c2} \approx 1.70$~K~\cite{Vollmer03,
Grube06, Measson08}. The coherence length of PrOs$_4$Sb$_{12}$ is $\xi=12$~nm, while muon spin rotation
($\mu$SR) measurements yield $\lambda(0) \approx 350$~nm~\cite{MacLaughlin02, Shu09}.

Sr$_2$RuO$_4$ and PrOs$_4$Sb$_{12}$ both have unusual superconducting states. Strong evidence for either line
nodes or deep gap minima, or both, in Sr$_2$RuO$_4$ comes from, {\it e.g.}, RF~\cite{Bonalde00},
microwave~\cite{Baker09} and specific heat measurements~\cite{Deguchi04A, Deguchi04B}. In PrOs$_4$Sb$_{12}$,
RF and thermal conductivity measurements suggest point nodes~\cite{Chia03}, while $\mu$SR and Sb NQR
show fully-gapped superconductivity (at ambient pressure)~\cite{MacLaughlin02, Shu09, Kotegawa03}. Nodes may
exist on a small-gap band suppressed by modest applied fields~\cite{Shu09}. Additional thermal conductivity
measurements confirm multi-band superconductivity, but indicate no nodes on either gap~\cite{Seyfarth06}.  In
Sr$_2$RuO$_4$, specific heat measured against applied field confirms multi-band
superconductivity~\cite{Deguchi04A}, and that if not nodes there are at least deep minima on the primary band.

Triplet pairing has been shown in Sr$_2$RuO$_4$ with high certainty by measurement of the $^{17}$O Knight
shift~\cite{Ishida98}. Measurement of the muon Knight shift in PrOs$_4$Sb$_{12}$ also indicates triplet
pairing, but with less certainty: owing to low-lying crystal electric field states, the expected Knight shift
for singlet pairing is less clear~\cite{Higemoto07}.

The absence of a Hebel-Slichter peak in Ru NMR measurements on Sr$_2$RuO$_4$~\cite{Ishida97}, and in Sb NQR
measurements on PrOs$_4$Sb$_{12}$~\cite{Kotegawa03, Kawasaki08} indicate unconventional superconductivity in
both materials. In Sr$_2$RuO$_4$ it is confirmed by demonstration that $T_c \rightarrow 0$ as the mean free
path shrinks to $\sim$$\xi_{ab}$~\cite{Mackenzie98}. In PrOs$_4$Sb$_{12}$, $T_{c1}$ shows a modest, and
$T_{c2}$ possibly a more pronounced, sensitivity to sample quality~\cite{Measson08}.  Odd-parity orbital
symmetry in Sr$_2$RuO$_4$ has been shown by fabrication of a $\pi$-SQUID~\cite{Nelson04}, and
superconductivity-related time-reversal symmetry-breaking (TRSB) by measurement of the Kerr
effect~\cite{Xia06}. A two-component order parameter is indicated by Josephson
interferometry~\cite{Kidwingira06}, hysteretic transport in microstructures~\cite{Kambara08}, and a jump in
the transverse sound velocity at $T_c$~\cite{Okuda03}. These results have all been interpreted in terms of a
chiral $p_x \pm ip_y$ orbital order parameter. The order parameter of PrOs$_4$Sb$_{12}$ remains an open
question; possibilities for both singlet and triplet pairing are listed in Ref.~\cite{Mukherjee06}.

Along with the shared features described above, Sr$_2$RuO$_4$ and PrOs$_4$Sb$_{12}$ are also 
the only two materials where observation of spontaneous,
superconductivity-related TRSB fields by $\mu$SR is well-established.  In Sr$_2$RuO$_4$, an average internal
induction far below $T_c$ of $\sim$0.5~G is found, with the rapid initial decay of muon polarization
indicating a peak induction of at least $\sim$5~G~\cite{Luke98, Luke00}, indicating a dilute density of
sources. In PrOs$_4$Sb$_{12}$, the average internal induction is at least twice as large,
$\sim$1.5~G~\cite{Aoki03}, but its distribution is not as peaked as in Sr$_2$RuO$_4$, suggesting a higher
source density. The TRSB appears to onset at the upper ($\sim$1.85~K) $T_c$.

For $p_x \pm ip_y$ order in Sr$_2$RuO$_4$ a magnetization along the $z$ axis (crystalline $c$ axis) is
expected: the orbital angular momentum of the condensate would give an uncancelled current at domain edges
(meaning domain walls and sample edges). Inward from these edges, Meissner screening would in turn result in
counterflowing screening
currents.  If each pair in the $\gamma$-sheet condensate is assigned angular momentum $\hbar$,
an edge current of $k_F^2 \hbar e / 8 \pi m^* = 2.6$~$\mu$A per layer results~\cite{Stone04}, for a field
discontinuity of 50~G. Matsumoto and Sigrist (MS) have solved the Bogoliubov-de Gennes equations in a
quasiclassical approximation for an ideal $p_x \pm ip_y$ superconductor (without secondary bands or 
gap minima/nodes, and with specular scattering at edges) and obtain edge and domain wall inductions
peaking at $\sim$10 and $\sim$20~G, respectively~\cite{Matsumoto99}. (We scale their unitless results by
$\xi=66$~nm and $\lambda_L=165$~nm; they assume $\kappa=2.5$.)

There is conflicting experimental guidance on domain size: Kerr rotation indicates domains at least
a few times larger than the beam size, or $\gtrsim$100~$\mu$m~\cite{Xia06}, while Josephson interferometry
suggests domains at the edges $\sim$1~$\mu$m across~\cite{Kidwingira06}. The $\pi$-SQUID required phase
coherence across the 0.6~mm width of the Sr$_2$RuO$_4$ crystal, indicating large domains~\cite{Nelson04}.  If
the MS result is approximately correct ({\it i.e.}, a $\sim$10~G field across a width $\sim$2$\lambda_{ab}$ at
domain walls), the 0.5~G average interal field observed by $\mu$SR suggests $\sim$10~$\mu$m domains.
Domain size might be affected by the cooling rate through $T_c$, which was $\sim$1~K/hr for the Kerr,
Josephson interferometry, and $\pi$-SQUID experiments, and faster for the $\mu$SR measurement. Domain size is
discussed in more detail in Ref.~\cite{Kallin09}.

To date, edge and domain wall fields have not been observed by scanning magnetic probes in Sr$_2$RuO$_4$.
Scanning Hall probe measurements by Bj\"{o}rnsson {\it et al} constrain edge and domain wall currents to be
less than $\sim$3 and 8\% of the MS results, repectively~\cite{Bjornsson05, Kirtley07} [The scan area was
(70~$\mu$m)$^2$, so domain walls may have been absent.] Kirtley {\it et al}, in SQUID scans spanning a
$\approx$1~mm-wide sample, improve the limit on both edge and domain wall currents to about 1\% of the
expectation, for domains larger than $\sim$8~$\mu$m~\cite{Kirtley07}. 

Analysis by Ashby and Kallin show that nonspecular or pair-breaking edge scattering could reduce the expected
edge magnetization, but not by the orders of magnitude required for consistency with experiment~\cite{Ashby09}.
Selection of Ginzberg-Landau parameters nearer the edge of stability for $p_x + ip_y$ order reduces the
edge currents, but also spreads them out over a larger range, so the total flux, and the observed signal in
scanning probes microns above the surface, would not be greatly reduced.

The work here further tightens the limits on chiral currents and domain structure in Sr$_2$RuO$_4$. We also
discuss the possibility of periodic domains, and show the first magnetic scans of PrOs$_4$Sb$_{12}$. 
\\

\noindent{\large \bf \sffamily Magnetic scans of Sr$_2$RuO$_4$}
\\

\begin{figure}[ptb]
\includegraphics[width=3.25in]{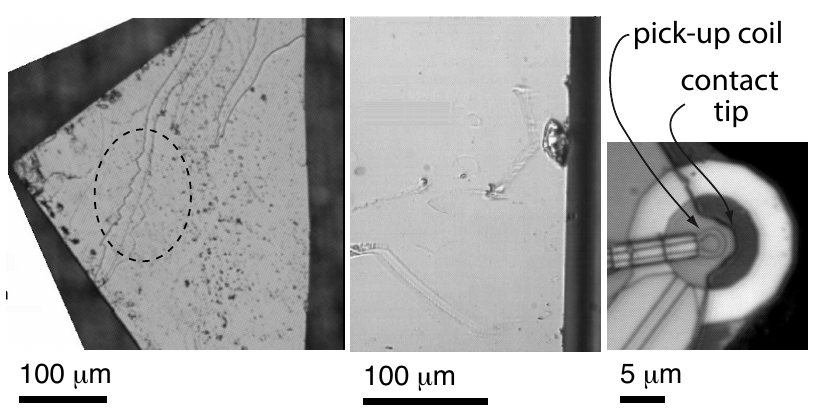}
\caption{\label{photos} 
Left: photograph of the Sr$_2$RuO$_4$ sample scanned in this work. The surface is a cleaved $ab$ surface. The
left edges are polished and the right is a growth edge. Cleave terraces visible in the magnetic scans
are circled.  Middle: PrOs$_4$Sb$_{12}$. The right edge is a growth edge. Right: front end of the SQUID used
in this work. The pick-up coil (of diameter 3.2~$\mu$m) and contact tip, the point where the SQUID contacts
the sample, are indicated. }
\end{figure}

The scanning SQUID used here is a niobium-based device, described in Refs.~\cite{Koshnick08}
and~\cite{Huber08}. Flux is coupled into the SQUID through a 3.2~$\mu$m-diameter pick-up coil (\Fig{photos}).
The leads to the pick-up coil are shielded to minimize flux coupling into the space between the leads. SQUIDs
are flux-sensitive devices, so the units on the data shown in this work are units of flux;
1~$\Phi_0 \equiv hc/2e$ in a 3.2~$\mu$m loop corresponds to an average induction of $\approx$2.5~G. 
The SQUID was generally scanned on a plane $\sim$1~$\mu$m out of contact with the sample, to reduce noise and
spurious features from surface roughness.

\begin{figure}[ptb]
\includegraphics[width=3.25in]{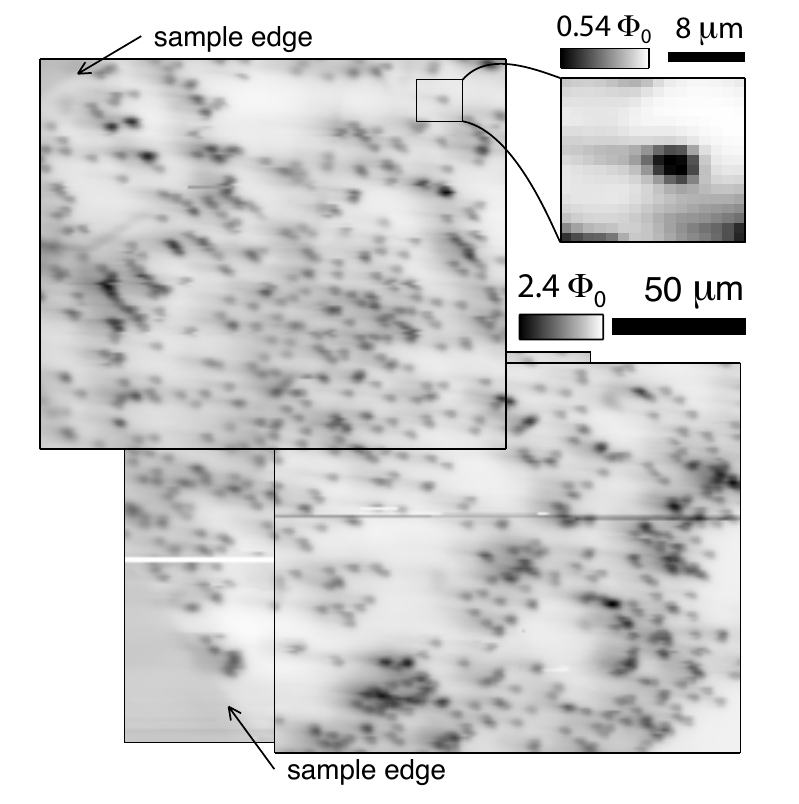}
\caption{\label{SRO_vortices}Three overlapping SQUID magnetometry scans of the Sr$_2$RuO$_4$ crystal pictured
in \Fig{photos}, in Earth's $\sim$1/2~Oe field. Single vortices are clearly visible. Across the entire scan
range vortices are lined up along upper-left-to-lower-right stripes.  A close-up of a single vortex is shown
at the upper right; the weak tail extending leftwards from the vortex is an artifact of the SQUID imaging
kernel. }
\end{figure}

\begin{figure}[ptb]
\includegraphics[width=3.25in]{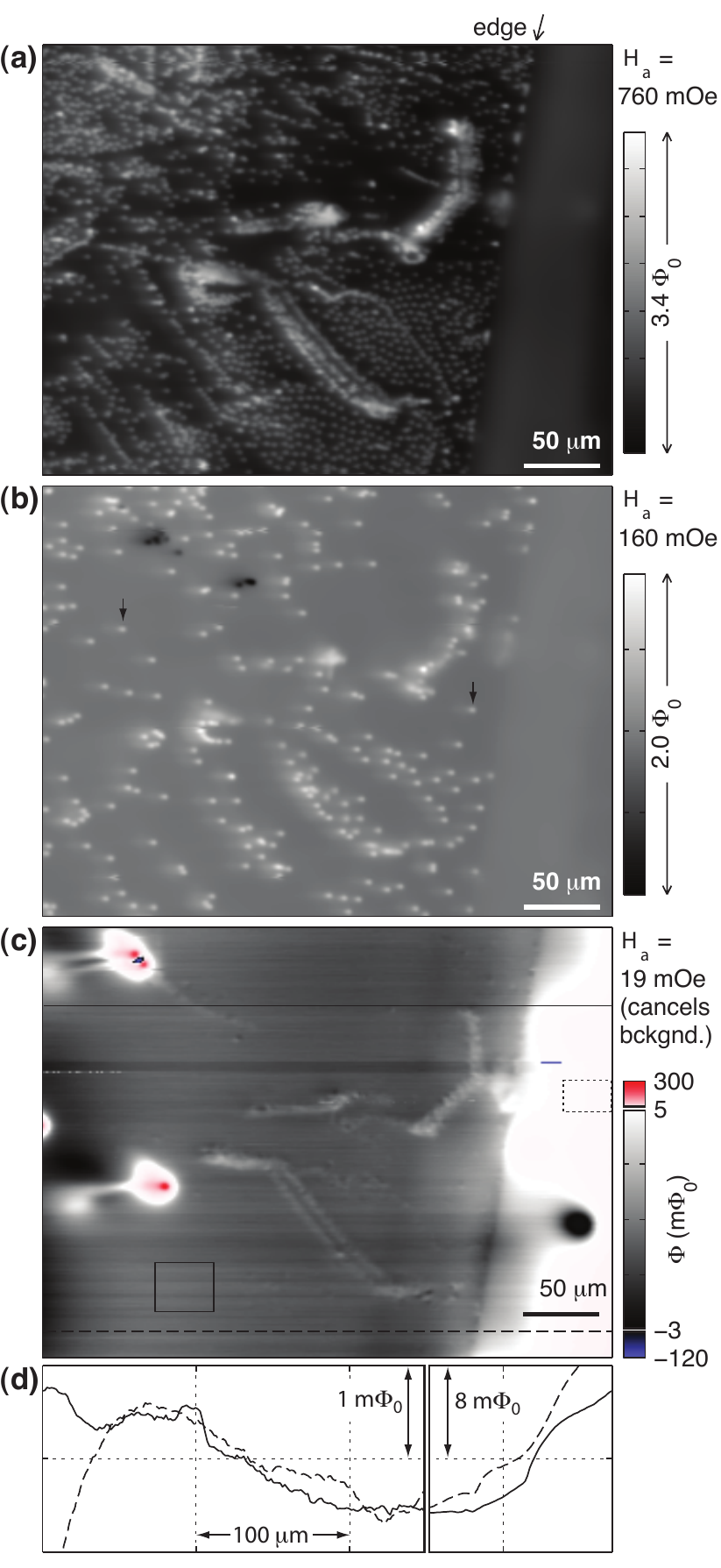}
\caption{\label{Pros_vortices} (a-c) Magnetic scans of the PrOs$_4$Sb$_{12}$ sample shown in \Fig{photos}. All
scans are at 0.4~K, under different cooling fields, $H_a$.
Vortices cluster at extended defects.
The indicated vortices in (b) are used for scan height determination.
In (c), $H_a=19$~mOe cancels the background $z$-axis field. Four vortices remain towards the left side of the
scan, and a fifth in a deep trench on the right side.
Note the broken colour
scale: the vortex signal greatly exceeds other features. Away from the vortices, surface
defects appear as an artifact of SQUID-sample interaction. A shadow $\sim$70~$\mu$m left of the edge
is an artifact of the imaging kernel and the edge. No magnetic features appear that could explain the
$\sim$1.5~G internal induction observed by $\mu$SR. After local plane subtraction, the rms signal in the solid
box is 11~m$\Phi_0$, and in the dotted box, 10~m$\Phi_0$. (d) Sections along the solid and dashed lines in
(c).}
\end{figure}

The Sr$_2$RuO$_4$ crystal (photograph: \Fig{photos}) was grown in a floating zone furnace, and was not
annealed after growth~\cite{Mao00}. All scans were done at $T=0.4$~K, with fairly rapid cooling through $T_c$,
$\sim$1~K/min.
For the mosaic of scans shown in \Fig{SRO_vortices}, the crystal
was cooled in Earth's $\sim$1/2~Oe field with the
out-of-plane component cancelled by $\sim$50\% by an applied field, however the applied field was subsequently
turned off and vortices reintroduced into the sample, to approximately match the ambient field, by
electromagnetic noise from the positioners. The weak tails that extend leftwards from each vortex, clearly
visible in the close-up of a single vortex, are artifacts of the imaging kernel: the
shielding of the pick-up coil leads is not perfect. The full extent of these tails can be seen in
\Fig{Pros_vortices}(c): $\approx$70~$\mu$m, about the distance between the pick-up coil and the point
where the shielding becomes a fully-formed coaxial cable.

Although not the main point of this paper, the mixed-state scans contain three features worth noting.
(1) The local vortex distribution is uneven. Even at small applied fields, where direct vortex-vortex
interaction is negligible, minimization of global field energy encourages a homogeneous vortex density---
compare with the much more homogeneous vortex distribution in clean areas of PrOs$_4$Sb$_{12}$, at an applied
field of 760~mOe [\Fig{Pros_vortices}(a)].
Local vortex coalescence has been previously reported in Sr$_2$RuO$_4$~\cite{Dolocan05, Hasselbach07,
Bjornsson05}, and also in MgB$_2$, where it was explained as originating from the
two-component order
parameter, one in the type-II regime, and the other type-I~\cite{Moshchalkov09}. As Sr$_2$RuO$_4$
is only weakly type-II ($\kappa_{ab} = \lambda_{ab}(0)/\xi_{ab}(0) \approx 2.6$; $\kappa>1/\sqrt{2}$ indicates
type-II) and has a two-component order parameter, the cause of local clustering might be similar. (2)
The large-scale distribution is also uneven: towards the lower right and
upper left, vortex-free areas up to $\sim$30~$\mu$m across are adjacent to similar-sized regions of
high vortex density. These regions may indicate spatially-varying sample quality.
(3) The distribution is anisotropic: over the
entire $\approx$300~$\mu$m-wide scan area, vortices line up along upper-left-to-lower-right stripes. A
striped vortex distribution was also reported in~\cite{Bjornsson05}. In highly anisotropic superconductors
$c$-axis vortices can form chains along in-plane vortices, a
phenomenon that has been observed in Sr$_2$RuO$_4$~\cite{Dolocan06}. However an in-plane field of
$\sim$10~Oe
appears to be required to form and orient well-defined chains~\cite{Dolocan05}, whereas the field for
\Fig{SRO_vortices} was no more than Earth's $\sim$1/2~Oe, and $\sim$50~mOe in Ref.~\cite{Bjornsson05}.
\begin{figure}[ptb]
\includegraphics[width=3.25in]{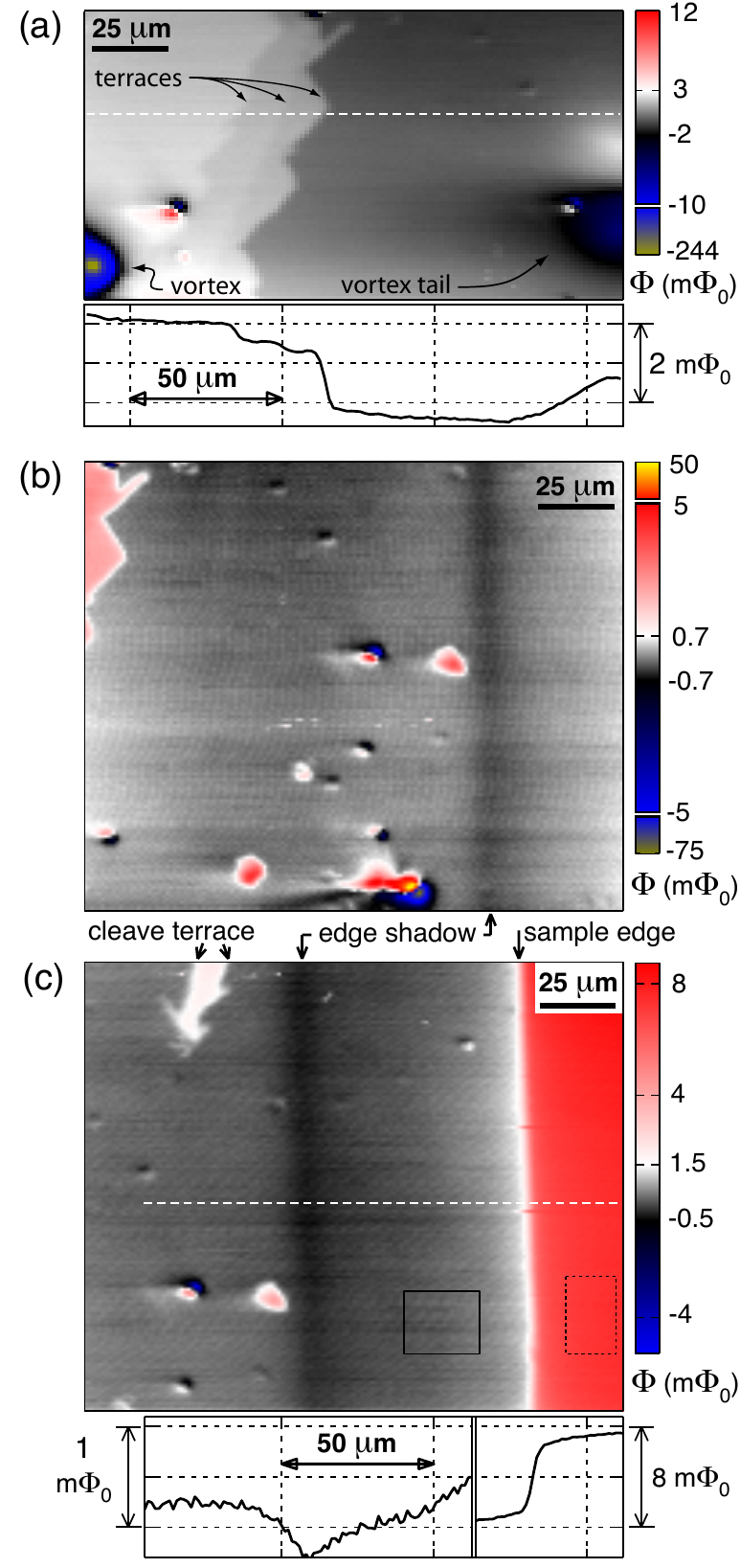}
\caption{\label{SRO_scans}Overlapping scans of Sr$_2$RuO$_4$ at 0.4~K; each is a separate thermal cycle.
In none of these three scans does the spontaneous edge and domain wall magnetization expected from chiral
superconductivity appear. (a) A scan over the cleave terraces, of total height $\sim$700~nm, indicated in
\Fig{photos}(a), and (lower panel) a section along the dashed line. Note the broken color scale: the peak
vortex signal far exceeds all other features. 
The terrace signal is an artifact of SQUID-sample interaction. The vortex tail, from a vortex
$\sim$50~$\mu$m beyond the right edge of the scan, is an imaging kernel artifact.
(b) An area with fewer terraces.
The ``edge shadow" is an imaging kernel artifact. Other features include dipoles (magnetic inclusions) and
some terraces. 
(c) A scan over the sample edge, and (lower panel) a section along the dashed line.
The rms signal in the solid and dotted boxes, after local plane subtraction, are
0.06~m$\Phi_0$ and 0.05~m$\Phi_0$, respectively. }
\end{figure}

\Fig{SRO_scans} shows this paper's main results on Sr$_2$RuO$_4$: magnetic scans in the near-absence of
vortices. The scans overlap, however each is a separate thermal cycle (to above $T_c$). Four 
prominent features appear in the scans: (1) In panel~(a), there are two vortices, one at the lower
left, and the other beyond the right edge of the scan. (The extended dipole-like feature is part of its tail.)
(2) The step changes in the signal across cleave terraces and the sample edge result from SQUID-sample
interaction, not static magnetization: the SQUID is scanned on a plane $\sim$1~$\mu$m above the sample
surface, so when the pick-up coil passes over topographic features its proximity to the sample changes.  In
operation, the SQUID is biased with a DC voltage and the current, which varies with the SQUID's critical
current, is measured.  Changing the proximity between the pick-up coil and a metallic surface affects the
inductance of that arm of the SQUID, which in turn affects the critical current (as does varying flux in the
SQUID, the desired signal)~\cite{Tesche77}.  (3) An edge shadow, the blurred dark line in panels~(b) and (c),
is also an artifact. It appears to correspond to a step change in magnetic field across the sample edge, with
the $\sim$70~$\mu$m distance between the shadow and the edge being set by the far extent of the SQUID imaging
kernel, {\it i.e.} the tails discussed above.  (4) There are numerous magnetic dipoles, most very weak:
compare their signals with that from the vortex. The source of the dipoles was not investigated,
however typical dipoles in the scans, with peak signals of 0.1--5~m$\Phi_0$, could be accounted for by
inclusions of SrRuO$_3$, a ferromagnet with 1.6--2~$\mu_B$/Ru~\cite{Dodge99}, $\sim$20-80~nm on a side. (Tails
on some of the dipoles, like the vortex tails, are imaging kernel artifacts.)
\\

\noindent{\large \bf \sffamily The absence of observed spontaneous magnetization in Sr$_2$RuO$_4$}
\\

No edge or domain wall magnetization, expected for $p_x \pm ip_y$ order, is apparent in the Sr$_2$RuO$_4$
scans. To compare this null result with theoretical expectation, it is necessary to account for the finite
sensor resolution and scan height. We model the pick-up coil as a 3.2~$\mu$m-diameter wire loop parallel to
the surface, with perfect coupling of $B_z$ inside and zero outside. Empirical scan heights consistent with
this model are obtained by studying the vortex and dipoles in \Fig{SRO_scans}.  For scan heights much larger
than $\lambda_{ab}$, the field distribution of a vortex approaches that of a monopole placed a depth
$\lambda_{ab}$ beneath the surface. The vortex in \Fig{SRO_scans}(a) has a FWHM (along $y$) of 3.8~$\mu$m and
a peak signal of $0.25\Phi_0$. In the monopole model, these values correspond to heights above the monopole of
1.6 and 1.9~$\mu$m, respectively, or, subtracting $\lambda$, $\approx$1.6~$\mu$m above the sample surface.
Heights above the dipoles are obtained by fitting the 2D scan data to the simulated response from a point-like
dipole, with dipole strength, dipole orientation and scan height as free parameters. The very prominent dipole
near the bottom of \Fig{SRO_scans}(b) gives a height of 1.7~$\mu$m. Fits to two other dipoles in
\Fig{SRO_scans}(b) give heights around 1.5~$\mu$m, and to two dipoles in \Fig{SRO_scans}(a), a few 0.1~$\mu$m
higher. In the Sr$_2$RuO$_4$ scans, the scan height above the surface is $z \approx 1.5$~$\mu$m.

\begin{figure}[ptb]
\includegraphics[width=3.2in]{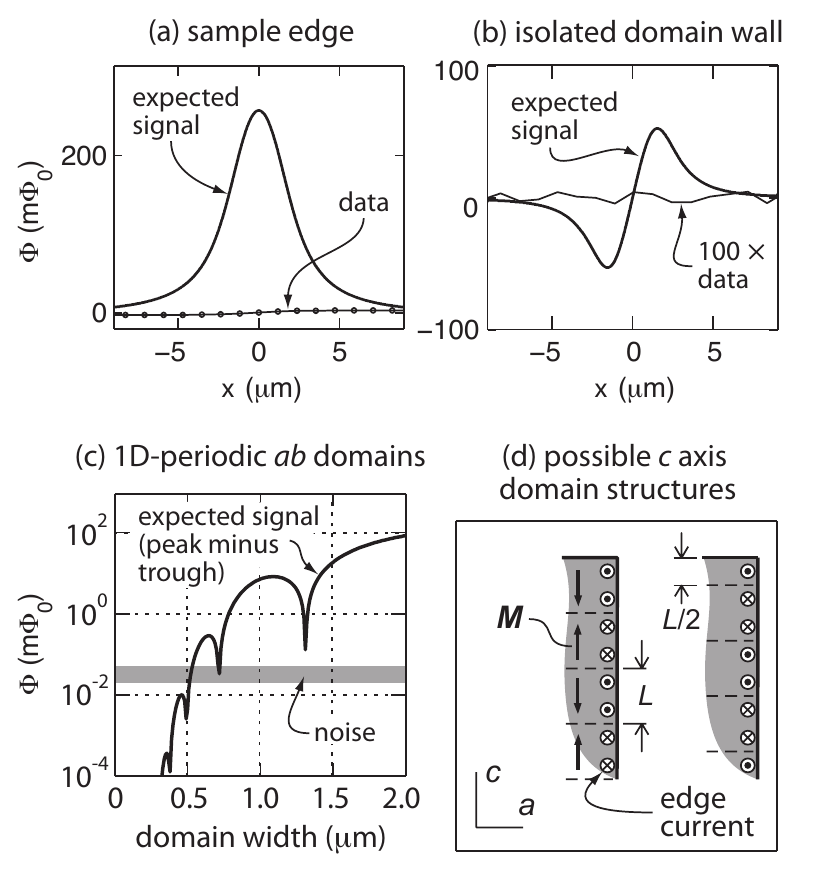}
\caption{\label{SRO_simulation}(a-c) Expected signals compared with observed data on Sr$_2$RuO$_4$ for (a) a
sample edge, (b) an isolated domain wall, and (c), domains periodic along one direction in the $ab$-plane. The
expected signals were obtained by extension of the MS results to a scan height of 1.5~$\mu$m, using eq.~1,
followed by averaging over a 3.2~$\mu$m-diameter pick-up coil. The data in (a) and (b) is along the dashed
line in \Fig{SRO_scans}(c), and the noise level indicated in (c) is the pixel-to-pixel noise in the boxes in
\Fig{SRO_scans}(c). (d) Hypothetical $c$-axis domain structures, with the magnetization direction and
equivalent edge currents indicated. Taking $|\mathbf{M}|=10$~G, for the cases at left and right our scans
indicate $L<20$ and 400~nm, respectively.}
\end{figure}

The MS calculation gives the magnetic induction deep beneath the upper surface of the sample. We extend their
results to the space above the sample following the procedure in Refs.~\cite{Kirtley07} and~\cite{Bluhm07}: If
the source inductions within the superconductor are along $z$, then \begin{equation} \tilde{B}_z(\mathbf{k},
z) = \frac{K}{k+K} \tilde{B}_{0,z}(\mathbf{k}) e^{-kz}, \end{equation} where $K=\sqrt{k^2 +
(1/\lambda_{ab})^2}$, $\tilde{B}_z(\mathbf{k})$ is the Fourier transform of $B_z(\mathbf{x})$, and $z$ is the
height above the sample surface.  The $K/(k+K)$ prefactor arises from Meissner screening: at $z \gg \lambda$,
its effect is that the field distribution at height $z$ is what it would be at $z + \lambda$ without the
prefactor.  We obtain the expected signal at edges and domain walls by applying eq.~1, with
$\lambda_{ab}=190$~nm, to the MS results, then averaging over a 3.2~$\mu$m-diameter circle.  (The accounting
of Meissner screening in eq.~1 is correct over featureless surfaces. However we apply eq.~1 to edges,
too: the error on $z$ exceeds $\lambda$, so this is a secondary error that does not affect our conclusions.)
Comparisons of expected and observed signals are shown in \Fig{SRO_simulation}. The expected signal is shown
for $z=1.5$~$\mu$m, however the choice of $z$ is not critical: a doubling of $z$ reduces expected edge signals
by $\sim$30\% and domain wall signals by $\sim$60\%.

\begin{figure}[ptb]
\includegraphics[width=3.2in]{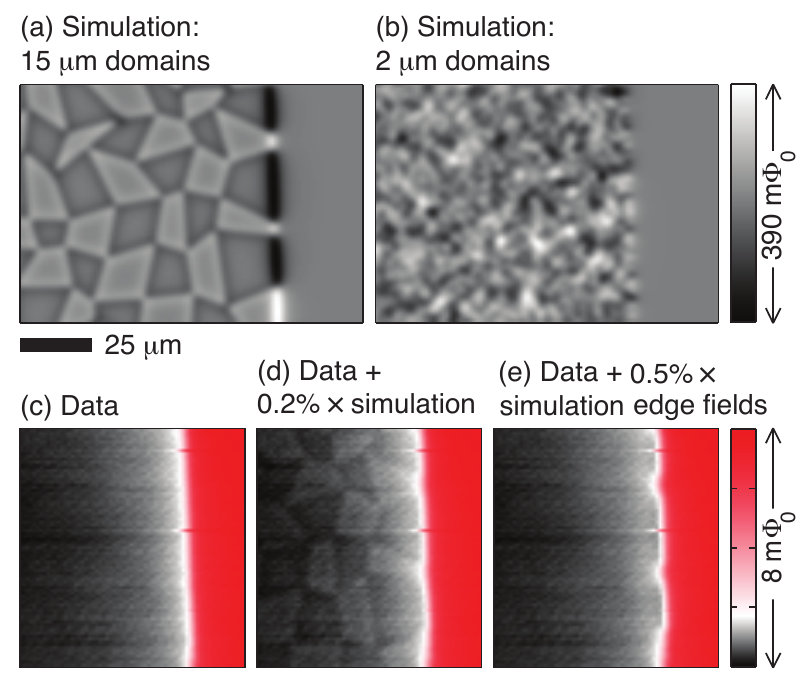}
\caption{ \label{SRO_simScans} (a,b) Simulated scans of a quasiperiodic domain structure in Sr$_2$RuO$_4$. The
MS results are extended to $z=2.0$~$\mu$m using eq.~1, then averaged over a 3.2~$\mu$m-diameter pick-up coil.
(c) Observed signal, from \Fig{SRO_scans}(c). (d,e) Observed signal plus a scale factor times the simulation.
In (e), domain wall contributions are removed.}
\end{figure}

\Fig{SRO_simScans}(a) and (b) show simulated scans, taking $z=2$~$\mu$m (the upper bound of the actual
scan height), for particular configurations of domains averaging 15 and 2~$\mu$m across, respectively. The
domains are quasiperiodic for computational convenience; their rms area variation is 38\% of the mean. In
panel (d) the simulated scans multiplied by a scale factor are added to the observed signal. For domains
2~$\mu$m and larger, domain wall fields would have been visible at $\sim$0.1\% of the MS result. To
look at edge fields separately, in (e) the edge fields alone of the simulation are added to the data. Weak
edge fields appear as modulations of the shape of the step, and would have been visible at $\sim$0.2\% of the
expectation for 4~$\mu$m or larger domains, or $\sim$1\% for 2~$\mu$m domains.

We cannot definitively exclude the possibility that the entire sample is a single domain,
and the edge signal is merged with the 6~m$\Phi_0$ step at the sample edge (due to
SQUID-sample interaction). In this case the limit on the edge signal is $\sim$6~m$\Phi_0$, or $\sim$3\% of the
expectation. However, we note again that the 0.5~G induction observed by $\mu$SR is a volume average: it is
highly unlikely that its source is entirely excluded from our $\sim$(150~$\mu$m)$^2$ scan area.

How large could domains be if they were completely random? For random domains magnetized along
$\mathbf{\hat{z}}$, the expected field at scan height $z$ is
given by eq.~15 of ref.~\cite{Bluhm07}:
\begin{equation}
\langle B_z^2 \rangle = \frac{15 \pi}{2} \frac{\lambda^3}{(z+\lambda)^6} M^2 V, 
\end{equation}
where $M$ is the typical magnetization and $V$ typical domain volume. $B_z$ would vary on a
length scale $\sim z$; here $z$ is comparable to the pick-up coil radius, so the signal is $\sim B_z z^2$. 
The observed signal over a featureless area of
the sample is 0.06~m$\Phi_0$~rms [see \Fig{SRO_scans}(c)], and over a section of vacuum,
0.05~m$\Phi_0$~rms. Taking $z=2$~$\mu$m, obtaining a signal less than 0.02~m$\Phi_0$ with $M=10$~[1]~G
requires $V<($6~nm)$^3$ [$<($30~nm)$^3$]. 

Larger domains are not excluded if they are periodic. \Fig{SRO_simulation}(c) shows the expected signal from
domains periodic along one direction in the $ab$ plane, taking the MS result for the domain wall
magnetization. Domains perfectly periodic along $a$ could have been as wide as $\sim$0.5~$\mu$m without being
detected.

Domains could also be periodic along $c$. To estimate a limit on $c$-axis domain widths, we take
$\mathbf{M}=(0,0,\pm M)$ everywhere, with $M$ constant; {\it i.e.} the magnetization is equivalent to thin
sheet currents at the sample edge. Meissner screening does not strongly affect the field distribution for
domain thicknesses $L \ll \lambda$, and so is neglected.  Two possibilities for $c$-axis periodic domains are
illustrated in \Fig{SRO_simulation}(d). If the domains are all of equal thickness $L$, then the upper half of
the top domain gives a far-field $|B|$ linear in $L$. (All remaining edge current is incorporated into
alternating dipoles.) If $|M|$ is 10~G, then for a 3.2~$\mu$m-diameter pick-up coil at heights $z=1$, 1.5 and
2~$\mu$m, $L=0.02$~$\mu$m gives peak edge signals of 2.5, 1.8 and 1.5~m$\Phi_0$, respectively. Alternatively,
if the thickness of the top domain is halved the long-range field scales as $L^2$.  For $L=0.40$~$\mu$m and
$z=1$, 1.5 and 2~$\mu$m, the peak edge signal is expected to be $\approx$3.8, 2.3 and 1.5~m$\Phi_0$ (in the
absence of Meissner screening, which would reduce the signal somewhat further).

A brief summary on the limits on chiral currents and domains sizes in Sr$_2$RuO$_4$:
\begin{enumerate}
\item For domains of $p_x \pm ip_y$ order larger than 2~$\mu$m across, domain wall currents are at most
$\sim0.1$\% of the expectation (as calculated by MS). Edge currents are at most $\sim$0.2\% of expectation for
4~$\mu$m and larger domains, and $\sim$1\% for 2~$\mu$m domains.
\item If domain wall currents are of the expected magnitude and domains are perfectly periodic stripes in the
$ab$ plane, the domain width can be up to $\sim$0.5~$\mu$m (periodicity 1~$\mu$m).
\item If $|\mathbf M| \sim 10$~G [1~G] and the domains are random, their volume is at most $\sim$(6~nm)$^3$
[$\sim$(30~nm)$^3$].
\item If $|\mathbf M| \sim 10$~G and $\mathbf{M}$ is periodic along $c$, the domain width can be up to $\sim
20$~nm if all domains are of the same width, or $\sim 400$~nm if the top domain is half the width of the
others, to give an edge signal $\lesssim1$~m$\Phi_0$.
\end{enumerate}
If a reason were found for chiral edge and domain wall currents to be vastly less than the MS result, then a
separate explanation for the source of the $\mu$SR signal would be required.
\\

\noindent{\large \bf \sffamily Magnetic scans of PrOs$_4$Sb$_{12}$}
\\

As with Sr$_2$RuO$_4$, in anticipation of possible edge currents an area of the PrOs$_4$Sb$_{12}$ extending
over an edge was selected for scanning. Scans of the mixed state under $H < 1$~Oe are shown in
\Fig{Pros_vortices}. [The edge is of a deep trench rather than the sample edge; a spread-out vortex, far
beneath the SQUID, appears at the lower right of \Fig{Pros_vortices}(c).] Also as with Sr$_2$RuO$_4$, the
cooling rate through $T_c$ was $\sim-1$~K/min.

The vortices clustered strongly at extended sample defects, the more prominent of which are visible in
the photograph. Where the sample is relatively free of defects [the lower right portion of
\Fig{Pros_vortices}(a)], the vortex distribution is homogeneous, in contrast to Sr$_2$RuO$_4$. 

A few opposite-sign vortices appear in panel (b), where the cooling field was 160~mOe, and in the same area
of a 300~mOe scan. They do not appear in panel (c), where the cooling field was near zero, and so are not
due to a magnetic inclusion.  They may be a consequence of trapping of positive-sign vortices combined with
overall flux expulsion: the vortex density in the 760~mOe scan approximately matches the cooling field, but
a significant portion of the flux was expelled in the 160~mOe scan (even after accounting for the nonzero
background field).
\begin{figure}[ptb]
\includegraphics[width=3.2in]{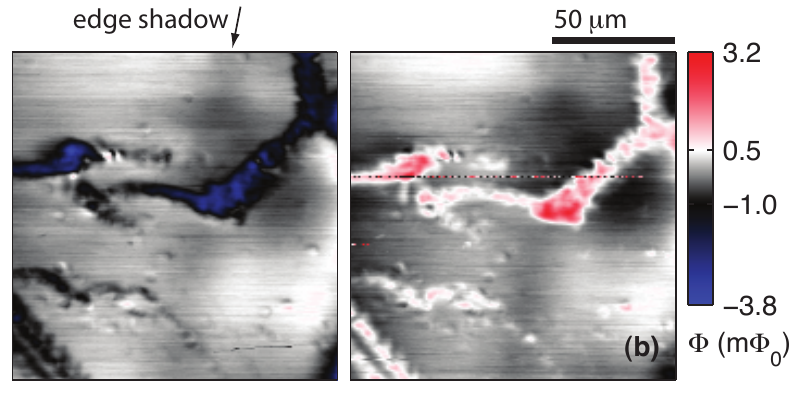}
\caption{\label{ProsReversingBias} Scans of the same area of PrOs$_4$Sb$_{12}$, with the SQUID bias current
reversed between the two scans. Features resulting from SQUID-sample interaction change sign, while features
resulting from static magnetic inductions do not.}
\end{figure}

Like the cleave terraces in Sr$_2$RuO$_4$, surface defects appear clearly in the nearly vortex-free scan
[\Fig{Pros_vortices}(c)], as a results of SQUID-sample interaction. This is proved by reversing the SQUID
bias: the surface defect signal changes sign whereas the edge shadow, resulting from static magnetic fields at
the edge, does not (\Fig{ProsReversingBias}).

The scan height over the PrOs$_4$Sb$_{12}$ sample can be determined by studying a few vortices, as in
Sr$_2$RuO$_4$. The isolated vortex in \Fig{Pros_vortices}(c) has a FWHM of 3.8~$\mu$m and a peak signal of
0.30$\Phi_0$, giving scan heights (above the monopole) in the monopole model of 1.6~$\mu$m and 1.5~$\mu$m,
respectively, or, subtracting $\lambda$, $z \approx 1.3$~$\mu$m. The indicated vortices in panel (b) yield
$z \approx 0.9$ and 1.0~$\mu$m, or $\approx$0.5~$\mu$m lower than in the Sr$_2$RuO$_4$ scans.

At a similar level to Sr$_2$RuO$_4$, no superconductivity-related TRSB fields are visible in the
PrOs$_4$Sb$_{12}$ scans.  Applying eq.~2, for the case of random magnetic domains, with $M=1$~G,
$\lambda=0.35$~$\mu$m and $z=1$~$\mu$m, a signal of less than 0.1~m$\Phi_0$ implies domains of volume less
than (30~nm)$^3$. The lower scan height allows a somewhat tighter limit on periodic domains than in
Sr$_2$RuO$_4$: if the domain magnetization is comparable to the MS result, the upper limit on domain width is
$\sim$0.4~$\mu$m.
\\

\noindent{\large \bf \sffamily Discussion and Conclusions}
\\

Ref.~\cite{Ashby09} discusses the possibility that domain wall fields in line with expectation account for the
$\mu$SR data, while edge fields are sharply reduced. Two mechanisms to achieve this are discussed:
pair-breaking edge scattering combined with a balancing of Ginzberg-Landau parameters, and a
competing edge order. If the induction observed by $\mu$SR is from domain walls, however, then as
discussed above it is unlikely that observable domain walls would have been absent from the scans here and in
Ref.~\cite{Kirtley07}. 
Mechanisms that
might eliminate edge {\it and} domain wall fields, while maintaining $p_x \pm ip_y$ order, include a
balancing of GL parameters such that the $p_x$ and $p_y$ components are everywhere nearly
equal~\cite{Logoboy09}, or the effects of multiple bands, or nodes / deep gap minima, or strong spin-orbit
coupling~\cite{Haverkort08}. However the origin of the TRSB fields observed in $\mu$SR would be left
unexplained.  Ref.~\cite{Aoki03} discusses, in regards to PrOs$_4$Sb$_{12}$, the possibility of a finite
hyperfine field induced at the $\mu^+$ sites, which would require a nonunitary order parameter. Nonunitary
pairing under $H=0$ has not been confirmed in any material, however, and is considered unlikely for
Sr$_2$RuO$_4$~\cite{SROreview}. 
Magnetization through pair-breaking at impurity sites is not likely to be the source of the $\mu$SR signal:
observable magnetization did not appear at defect sites, of which there must have been very many, across a
$\sim$100~$\mu$m range, in the scans here. Also, $\mu$SR data on a lower-$T_c$ sample, with more defects,
indicate weaker, not stronger, TRSB fields~\cite{Luke00}.

As described above, random static domains of magnetization consistent with the $\mu$SR data would need to
be on a scale $\sim \xi$ or smaller to have evaded detection. It seems unlikely that domains of an orbital
order parameter could be so small. Also, in Sr$_2$RuO$_4$, the field distribution observed by $\mu$SR is more
consistent with dilute sources than dense random magnetization. 

The limits on periodic domains are less severe. If edge currents are of the predicted order (accounting for
the $\mu$SR data), then the most plausible scenario described here for eliminating long-range edge fields may
be domain periodicity along the $c$-axis. If the periodicity is good then the domains could be up to
$\sim$400~nm $\sim$100$\xi_c$ in width, implying an energy cost below 1\% of the condensation energy.

Closely-spaced $c$-axis domain walls extending throughout the sample are not ruled out by experiments to date:
if the structure of a domain wall is such that one of $k_x$ or $k_y$ remains finite across the wall,
then by slow cooling it may be possible to obtain one of $k_x$ or $k_y$ finite throughout the crystal
--- domain formation may be controlled by dynamics near $T_c$, where $\xi$ is longer and domain walls may
interact more strongly --- which would permit fabrication of $\pi$-SQUIDs~\cite{Nelson04}. Also, supposing
that $k_x$ is the dominant component, a natural dichotomy, as observed~\cite{Kidwingira06}, would arise in
Josephson interferometry, between junctions along the face $\perp \mathbf{\hat{x}}$, where the phase would be
constant across the face, and $\perp \mathbf{\hat{y}}$, where it would alternate on a tight length scale.

However the energetics that would drive $c$-axis domain formation are not clear: Meissner screening of the
$k_x \pm ik_y$ edge currents cancels the long-range fields that encourage domain formation in conventional
ferromagnets. Also, the absence of magnetization at microscopic surface features, such as cleave terraces,
implies a tighter limit on $c$-axis domain width than $\sim$400~nm, such that $c$-axis domains should be
considered a possibility rather than a likelihood.

There is less experimental guidance on the possibilities for PrOs$_4$Sb$_{12}$. Because PrOs$_4$Sb$_{12}$ is
cubic, if the orbital OP turns out to break time reversal symmetry then a more complex domain structure than
in Sr$_2$RuO$_4$ is expected~\cite{AbuAlrub09}. {\it E.g.} in Ref.~\cite{Mukherjee06}, a singlet OP
$\sim(d_{yz} + id_{xz} + 0d_{xy})$ is illustrated, which would give six types of domains: $\pm x$, $\pm y$ and
$\pm z$. If a domain structure does exist in PrOs$_4$Sb$_{12}$, then the $\mu$SR data indicate that the
domains would likely be smaller (as scaled by $\lambda$) than in Sr$_2$RuO$_4$~\cite{Aoki03}.

In conclusion, magnetic scans of Sr$_2$RuO$_4$ and PrOs$_4$Sb$_{12}$ have been presented. No static
magnetization remotely near the TRSB fields observed by $\mu$SR in both materials, or theoretical expectation
for $p_x \pm ip_y$ order in Sr$_2$RuO$_4$, was observed.
\\

\noindent{\large \bf \sffamily Acknowledgements}
\\

This work was funded by the United States Department of Energy (DE-AC02-76SF00515). YM acknowledges funding
from a Grant-in-Aid for Global COE programs from MEXT, Japan. MBM acknowledges funding from the U.S.
Department of Energy. We express gratitude to Catherine Kallin, Daniel Agterberg, Manfred Sigrist, Graeme Luke
and Srinivas Raghu for useful discussions.

\end{document}